\documentclass[twocolumn,prl,aps,superscriptaddress,floatfix,showpacs]{revtex4}

\usepackage{graphicx}
\usepackage{multirow}

\begin{document}
\title{Metal-ligand interplay in strongly-correlated oxides:a parametrized phase diagram for pressure induced spin transitions} 
\author{A. Mattila} 
\affiliation{Division of X-Ray Physics, Department of Physical Sciences, POB 64, 00014 University of Helsinki, Finland} 
\author{J.-P. Rueff} 
\affiliation{Laboratoire de Chimie Physique -- Mati\`{e}re et Rayonnement, UMR 7619, 11 rue Pierre et Marie Curie, F-75231 Paris, France}
\affiliation{Synchrotron SOLEIL, L'Orme des Merisiers, BP-48 Saint-Aubin, 91192 Gif-sur-Yvette, France}
\author{J. Badro} 
\affiliation{Institut de Physique du Globe de Paris, F-75252 Paris, France} 
\affiliation{Institut de Min\'{e}ralogie et Physique des Milieux Condens\'{e}s, Place Jussieu, 75005 Paris, France}
\author{G. Vank\'{o}} 
\affiliation{European Synchrotron Radiation Facility, BP 220, F-38043 Grenoble, France} 
\affiliation{KFKI Research Institute for Particle and Nuclear Physics, P.O. Box 49, H-1525 Budapest, Hungary}
\author{A. Shukla} 
\affiliation{Institut de Min\'{e}ralogie et Physique des Milieux Condens\'{e}s, Place Jussieu, 75005 Paris, France}
\date{\today}   

\begin{abstract} 
We investigate the magnetic properties of archetypal transition-metal oxides MnO, FeO, CoO and NiO under very high pressure by x-ray emission spectroscopy at the K$\beta$ line. We observe a strong modification of the magnetism in the megabar range in all the samples except NiO. The results are analyzed within a multiplet approach including charge-transfer effects. The pressure dependence of the emission line is well accounted for by changes of the ligand field acting on the $d$ electrons and allows us to extract parameters like local $d$-hybridization strength, O-$2p$ bandwidth and ionic crystal field across the magnetic transition. This approach allows a first-hand insight into the mechanism of the pressure induced spin transition. 
\end{abstract}  
\maketitle  

Pressure is an effective means to drive the electronic density of a system, and thereby the electron interaction and delocalization. In correlated materials such as transition metal oxides (henceforth MO), lattice compression strongly impacts on spin and orbital degrees of freedom, while substantially affecting the transport, structural and magnetic properties. This sensitivity results from the considerable changes that pressure induces in the  internal energy of the system. Metal-insulator transitions or magnetic collapse are clear-cut illustrations of the effects of pressure through its influence on electronic correlation, charge transfer and magnetic stability. 
A satisfactory description of $d$ electrons in transition metal oxides constitutes, in fact, an ongoing challenge for theory. This is all the more relevant for pressure-induced phenomena like magnetic collapse which need proper treatment of $d$ hybridization. First-principle calculations can be cited in this context: high-pressure magnetism of MO has been treated in the GGA approximation~\cite{Cohen1997} by considering $d$ electrons as purely band-like under pressure. Correlation effects were more recently included in the treatment of high-pressure structural phases of FeO (LDA+U)~\cite{Gramsch2003}, FeSiO$_4$~\cite{Cococcioni2003,Jiang2004} and Fe$_2$O$_3$~\cite{Rollmann2004} (GGA+U) and spin-state in (Mg,Fe)O solid solution~\cite{Tsuchiya2006}.

Recent advances in high-pressure spectroscopy and improvements in theoretical modeling of strongly correlated systems are now in a position to further current understanding. We report here the results of x-ray emission spectroscopy (XES) at the K$\beta$ line in a series of prototypical transition-metal oxides, MnO, FeO, CoO, and NiO under very high pressure conditions. XES is now well established as a local probe of the transition metal magnetism. It is an all-photon technique, fully compatible with high-pressure sample environment, and relevant to the complete transition metal series. The data are analyzed in the light of multiplet calculations within the Anderson impurity model~\cite{Kotani2001,Groot2001,Groot1994_2}. Here, in contrast to band-like treatments of $d$ electrons, crystal-field, ligand bandwidth, and charge transfer are explicit parameters providing a physically intuitive insight. The model, derived from the configuration interaction approach, was first introduced to explain the core-photoemission spectra of transition metals~\cite{Zaanen1986,Mizokawa1994}. It was later applied to the K$\beta$ emission line in Ni-compounds~\cite{Groot1994} and more recently in transition metal oxides~\cite{Tyson1999,Glatzel2001,Glatzel2004}. The multiplet calculation scheme yields an accurate model of the emission lineshape and allows a direct estimate of the fundamental parameters through constraints imposed by comparison with the experimental lineshape. In this letter the method is applied to pressure induced magnetic collapse, giving the opportunity to study the given systems in two contrasting states from the electronic, magnetic and even structural point of view. This allows us to establish the frontiers of a phase diagram characterising this transition and sheds new light on the intermediate pressure regime.

\begin{figure}[htb] 
\includegraphics[width=7 cm]{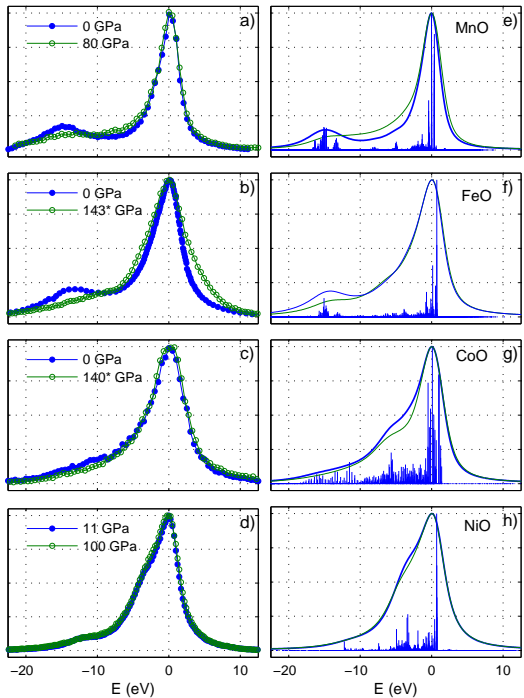}
\caption{\label{kbeta4} (color online) (a--d) K$\beta$-XES spectra measured in MnO, CoO, FeO and NiO in both low (close circles) and high pressure (open circles) phases. (*) indicates spectra obtained after laser-heating. (e--h) Calculated spectra at high (tick) and low pressures (thin lines). Ticks are raw low-pressure data before broadening. The spectra are normalized to the peak height.}
\end{figure}

We have measured XES at the K$\beta$ line in MnO, FeO, CoO and NiO at the European Synchrotron Radiation Facility using diamond anvil cells and Be gasket~\cite{Rueff2005,Badro2002}. The high pressure spectrum of FeO shown in panel~\ref{kbeta4}(b) has been measured at the Advanced Photon Source (using the setup described in Ref.\ \onlinecite{Badro2002}) after laser-heating at 140 GPa. The high pressure spectrum of CoO has been obtained after laser heating at high pressure. Fig.~\ref{kbeta4}(a--d) summarizes the K$\beta$ emission spectra measured in the transition-metal oxide series at low and high pressures. The spectra are aligned to the main peak energy and normalized to unity. They all present a satellite structure (known as K$\beta'$) peaking on the low energy side of the main emission line (K$\beta_{1,3}$). At high pressure, all the spectra but NiO show significant modifications in the lineshape, essentially observed in the satellite region. The asymmetric broadening of the main line in FeO at 140 GPa is an artifact.

Following the the treatment of Peng \textit{et al.}~\cite{Peng1994a} within the atomic multiplet formalism, the main line and satellite result from the exchange interaction between the $3p$ core hole and the $3d$ orbital in the final state, which splits the final state manifold. As a rule of thumb, the satellite is therefore expected to shrink with decreasing $3d$ magnetic moment and to move closer to the main peak. This agrees well with the observed spectral changes when going from MnO ($3d^5$, $S$=5/2) to NiO ($3d^8$, $S$=1). It also qualitatively accounts for the collapse of the satellite at high-pressure observed in MnO, FeO and CoO, viewed as the signature of the HS to LS transition on the given metal ion. The nearly identical spectra in NiO, where no such transition is expected, confirms the rule. 

The strong modification of the electronic properties at high pressure which accompanies the magnetic collapse, is usually accompanied by a structural transition. Which of the structural or magnetic transition initiates the other is not under discussion here since we do not discuss structural changes under pressure. 
But the tight correlation between the two phenomena remains, and can serve to fingerprint the phase transition at high pressure. In MnO, the HS-LS magnetic transition evokes the structural transition reported from the anti-ferromagnetic distorted rocksalt structure (AF-rB1 phase) to an intermediate phase around 80--90 GPa~\cite{Fang1999,Kondo2000} in the rB1 (or mixed rB1 and B8) structure. Above 105--120 GPa, another structural transition was observed which supposedly drives MnO to an hexagonal B8 structure. The latter transition is identified with a Mott transition~\cite{Patterson2004}. FeO presents a similar succession of structural changes from AF-rB1 to the metallic B8~\cite{Fei1994} or eventually to the inverse-B8 structures~\cite{Fang1999}. The FeO phase transition takes place around 90-100 GPa but is not fully achieved before $\sim$140 GPa~\cite{Pasternak1999} and heating of the sample to overcome kinetic barriers, in good agreement with the HS-LS transition. In CoO, the occurrence of a magnetic collapse around 100 GPa, which is associated with a lower atomic volume, agrees well with the phase transition toward a high-density rocksalt structure, observed around 90 GPa~\cite{Guo2002a}. Finally, the absence of a magnetic transition in NiO is supported by the stability of the rB1 structure, which has been reported up to 140 GPa~\cite{Eto2000}. 

The atomic description of XES, however, omits the crucial role played by the O($2p$)-M($3d$) charge-transfer effects and finite O-$2p$ bandwidth. Inclusion of the charge-transfer in the multiplet calculations also substantially improves the simulated main-peak to satellite intensity ratio, via a transfer of spectral weight to the main peak~\cite{Groot1994}. In the cluster model charge-transfer enters the calculations through a configuration interaction scheme within the single impurity Anderson model. We considered a linear combination of $3d^n$, $3d^{n+1}\underline{L}$ and $3d^{n+2}\underline{L}^2$ configurations, where $\underline{L}$ denotes a hole in the O-$2p$ state, for CoO ($n=7$) and NiO ($n=8$) ground states. Two configurations $3d^n$ and $3d^{n+1}\underline{L}$ were used for MnO ($n=5$) and FeO ($n=6$) \footnote{The inclusion of the third configuration should mainly decrease the estimated charge transfer energies (K. Okada and A. Kotani, J. Phys. Soc. Jpn. {\bf 61}, 4619 (1992)). This does not greatly affect the analysis, since we are not directly comparing the exact parameter values between different oxides.}. The calculations were made in the $O_h$ basis set at 300 K. The Slater integrals and spin-orbit parameters were calculated using the Hartree-Fock method~\cite{Cowan1981}, and the Slater integrals were further scaled down to 80\% to account for the intra-atomic configuration interaction effects. Crystal field effects were considered using the approach developed by Butler~\cite{Butler1981} and charge-transfer effects using the code by Thole and Ogasawara~\cite{Groot2001}. The K$\beta$ emission spectra were calculated taking into account the term-dependence of the final state lifetime broadening~\cite{Taguchi1997}.  

The model parameters were first chosen to reproduce the emission spectra at ambient pressure. The parameters, charge-transfer energy $\Delta$ (defined as the energy difference between the centers of gravities of $3d^n$ and $3d^{n+1}\underline{L}$ configurations), hybridization strength in the ground state $V_{e_{g}}$, the O-$2p$ bandwidth $W$, and the crystal field splitting $10Dq$ are summarized in table~\ref{param}. We used the same core hole Coulomb interaction $U_{dc}$ for both $1\underline{s}$ and $3\underline{p}$ core hole states. The hybridization strength for $t_{2g}$ symmetry states $V_{t_{2g}}$ was set to half of the value for $e_g$ states $V_{e_{g}}$ and for core hole states $V_{e_{g}}$ was reduced by 0.4~eV from the ground state value. The term-dependence of the final state lifetime broadening was approximated using the same linear dependence ($-0.2\times \omega$ full width at half maximum) of the broadening on the fluorescence emission energy $\omega$ for all the MO emission spectra. The spectra were finally convoluted with a 1.5~eV full width at half maximum Gaussian to account for the instrumental broadening.
For the high pressure emission spectra we first estimated the changes in the oxygen bandwidth $W$, hybridization $V_{e_{g}}$, and crystal field splitting $10Dq$ from the ambient pressure values using the dependence of the parameters on the oxygen distance~\cite{Harrison1980}. Finally, $\Delta$ was chosen to fit the experiment while allowing small variations of $10Dq$, $V_{e_{g}}$, and $W$. The final values are given in table~\ref{param}. To improve the fitting in CoO and NiO, the on-site Coulomb interaction $U$ was also included, yielding values of 6.0~eV and 9.2~eV, respectively. The core hole potentials and Slater integrals were kept at ambient pressure values. We approximated the local symmetry of the metal ion at high pressures with $O_h$ to reduce the number of model parameters, although some deviations from the octahedral symmetry occurs in the MO high pressure crystal phases.
Fig.~\ref{kbeta4}(e--h) shows the calculated spectra for both the ambient and high pressure phases. The degree to which the experimental data can be reproduced is remarkable considering the simplicity of the model used. This assures us that the parameters singled out by our approach are the relevant ones for the description of magnetic collapse. For MnO, CoO and FeO the calculations for the high pressure phases yield a LS ground state and for ambient pressures a HS ground state. For NiO with $3d^{8}$ configuration, no spin transition occurs in $O_{h}$ local symmetry. 
\begingroup
\squeezetable
\begin{table}
\caption{Parameters used in the calculations (in eV). The ground state configuration (HS/LS) is given in parenthesis after the pressure value. For NiO with $3d^{8}$ configuration, no spin transition will occur in $O_{h}$ local symmetry. Charge transfer energy is given by $\Delta$; $V_{e_{g}}$ is the hybridization strength; $10Dq$ the crystal-field splitting; $U_{dc}$ the core hole Coulomb interaction; and $W$ denotes the oxygen $2p$ bandwidth. }
\label{param}
\begin{ruledtabular}
\begin{tabular}{lcccccc}
  & P (GPa) & $\Delta$ & $V_{e_{g}}$ & $10Dq$ & $U_{dc}$ & $W$(O-$2p$) \\
\hline
\multirow{3}{*}{MnO} & 0 (HS) & 5.0 & 2.2 &  1 & 10.0 & 3.0 \\
 & 80 (LS) & 6.0 & 3.06 &  1.6 & 10.0 & 4.0 \\
 & 100 (LS) & 6.0 & 3.7 & 2.3 & 10.0 & 6.0 \\
\hline
\multirow{2}{*}{FeO} & 0 (HS) & 5.0 & 2.4  & 0.5 & 7.0 &5.0  \\
& 140 (LS) & 5.0 & 3.2  & 0.8 & 7.0 & 9.0  \\
\hline
\multirow{2}{*}{CoO} & 0 (HS) & 6.5 &  2.5 & 0.7 & 7.0 & 4.0  \\
& 140 (LS) & 6.5 &  4.2 & 1.2 & 7.0 & 8.0 \\
\hline
\multirow{2}{*}{NiO} & 0  & 3.5 & 2.4  & 0.3  & 9.0 & 5.0  \\
& 100 & 4.5 & 3  & 0.65 & 9.0 & 7.5  \\
\end{tabular}
\end{ruledtabular}
\end{table}
\endgroup

\begin{figure}[htb] 
\includegraphics[width=0.95\columnwidth]{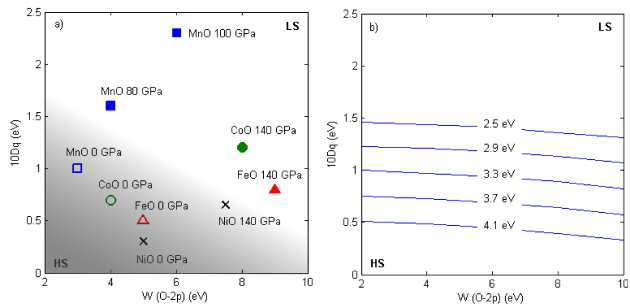}
\caption{\label{phase_diag} (color online) Phase diagram of the magnetic collapse in the transition-metal oxides. (a) The symbol coordinates refer to calculated values of $10Dq$ and $W$(O-$2p$), for the HS (open symbols) and LS (closed symbols) states. Crosses indicate the absence of spin transition. The shaded area is a guide to the eyes separating the two regions. (b) The lines mark the calculated HS-LS transition boundary for $d^{7}$ configuration for different values of $V_{e_{g}}$ shown in the figure. Charge-transfer energy $\Delta$ was set to 6.5~eV.} 
\end{figure}

Our findings, relating parameter values to the spin state of the system are summarized in Fig.~\ref{phase_diag}. The role of these parameters taken separately has been emphasized in earlier approaches, that of the crystal-field splitting $10Dq$ being paramount in the atomic description of XES and relevant for dilute systems while band-like calculations obviously take into account the O-$2p$ bandwidth $W$. Here we explicitly show the HS-LS transition as resulting from the conjugated effects of increase of the crystal-field splitting $10Dq$ \emph{and} a broadening of the O-$2p$ bandwidth $W$ together with the covalent contribution from the hybridization to the ligand field~\cite{Groot1994_2} at high pressures. While the increase of $10Dq$ clearly drives the system towards a LS state our analysis nicely highlights the interplay of the ligand bandwidth together with the hybridization. Fig.~\ref{phase_diag}(a) shows the calculated values of $10Dq$ and $W$ for the four oxides across the magnetic transition. Both quantities increase to produce magnetic collapse though in varying proportions according to the system in question. This variation allows us to establish a rough frontier between the HS and LS states as shown by the difference in shading in the figure. The sizeable increase of $10Dq$ in MnO with pressure contrasts with the other oxides where it is smaller, the essential increase being in the oxygen bandwidth $W$(cf.\ Fig.~\ref{phase_diag}(a)), which would tend to drive the oxides towards a metallic ground state~\cite{Shukla2003}. This possibly explains the larger volume compressibility of MnO compared to FeO, CoO and NiO~\cite{Fang1999,Kondo2000,Guo2002a,Eto2000}, and the high pressure transition towards the $B8$ compact structure. 
The effect of simultaneous evolution of these parameters together with hybridization across the magnetic collapse transition is represented in Fig.~\ref{phase_diag}(b). The lines mark the calculated HS-LS transition boundary for the different values of $V_{e_{g}}$ for a $d^{7}$ configuration. The increase of bandwidth $W$ contributes to the stabilization of the LS ground state in addition to the effect of hybridization.

\begin{figure}[htb] 
\includegraphics[width=5.5 cm]{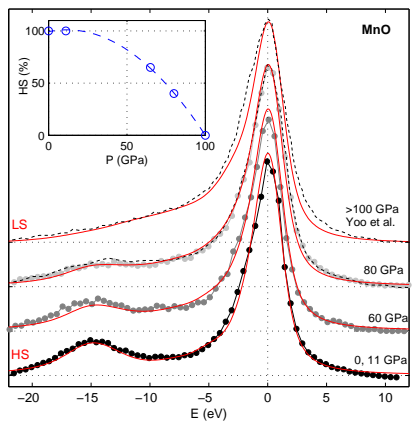}
\caption{\label{mno} (color online) XES spectra in MnO (circles) as a function of pressure. Data from Ref.\ \onlinecite{yoo2005} are shown with dashed lines at 80 and 100 GPa. Solid (red) lines are the calculated spectra. The intermediate pressure points have obtained from a linear combination of the HS and LS (100 GPa) spectra. The amount of spin-state is indicated in the inset; the dashed-line is a guide to the eyes.}
\end{figure}

We can ask ourselves if this HS/LS transition is abrupt and to what degree the frontier between the two regions is well marked. As an example we show in Fig.~\ref{mno} the emission line in MnO at four pressure points across the magnetic transition (circles). In addition to our dataset (0, 60 and 80 GPa), spectra from Ref.~\onlinecite{yoo2005} measured at 80 GPa (to verify consistency) and 100 GPa (dashed-line) were used. The calculated spectra (solid line) at intermediate pressures were fitted to the experimental spectra using a linear combination of the theoretical HS and LS (100 GPa) spectra. The amount of HS state derived from the fit is indicated in the inset. The Mn spin state gradually decreases from ambient to the high pressure, supposedly reaching a full LS state around 100 GPa which coincides with the reported Mott pressure transition. The predicted moment in the metallic phase is $\sim$1$\mu_B$~\cite{Cohen1997} in good accordance with a pure LS configuration. Thus the transition is characterized by an intermediate regime where a mixed quantum-state exists with both HS and LS states coexisting \cite{Haverkort2006}.  
Interestingly enough, the computed energy separation between the HS and LS multiplets in MnO at 80 GPa is 12~meV, indicating that the HS state can be thermally populated although it does not constitute the true ground state. As pressure is increased, the population of the two states is progressively reversed, leading to full LS conversion eventually. Increase in temperature would further broaden the pressure range of this intermediate regime~\cite{Tsuchiya2006,Sturhahn2005}. In other words, the sluggishness of the magnetic transition potentially results from a homogeneous superposition of HS and LS states. The sluggish transition is a widely observed feature in transition-metal compounds, both in the magnetic (e.g. see~\cite{Pasternak1999,Rozenberg2003}) or structural data. We further suggest that the sluggishness of the electronic transition may be enhanced through the interplay with the ligand as seen in Fig.~\ref{phase_diag}(b), where large changes in $W$ are needed to bring about changes in the magnetic state. In pure Fe on the contrary a sharp transition is observed~\cite{Mathon2004}. 

Aleksi Mattila has been supported by the Academy of Finland (Contract No. 201291/205967/110571). We thank the GSECARS beamline staff at APS for help during the FeO experiment and Akio Kotani for discussions.

\end{document}